\documentclass{aa}
\usepackage{graphicx}
\usepackage{txfonts}
\usepackage{natbib}
\usepackage{rotating}

\begin{document}
\title{Identification of blue high proper motion objects in the Tycho-2 and 2MASS catalogues using Virtual Observatory tools}
\titlerunning{Blue high proper motion objects in Tycho-2 and 2MASS}
%
%
\author{
F. M. Jim\'enez-Esteban\inst{1,2,3}
\and
J. A. Caballero\inst{4}
\and
E. Solano\inst{1,2}}
\institute{
Centro de Astrobiolog\'{\i}a (INTA-CSIC), Departamento de Astrof\'{\i}sica, PO
Box 78, E-28691, Villanueva de la Ca\~nada, Madrid, Spain\\ 
\email{fran.jimenez-esteban@cab.inta-csic.es}
\and
Spanish Virtual Observatory
\and
Saint Louis University, Madrid Campus, Division of Science and Engineering,
Avenida~del~Valle 34, E-28003 Madrid, Spain
\and
Centro de Astrobiolog\'{\i}a (INTA-CSIC), Departamento de Astrof\'{\i}sica,
Carretera de Ajalvir km~4, E-28850 Torrej\'on de Ardoz, Spain} 
\date{Received 16 Jun 2010 / Accepted 12 Sep 2010 }

\abstract
{}
{With available Virtual Observatory tools, we looked for new bright
  blue high proper motion objects in the entire sky: white dwarfs, hot
  subdwarfs, runaway OB stars, and early-type stars in nearby young
  moving groups.}
{We performed an all-sky cross-match between the optical Tycho-2 and
  near-infrared 2MASS catalogues with Aladin, and selected objects
  with proper motions $\mu$\,$>$\,50\,mas\,yr$^{-1}$ and colours
  $V_T-K_{\rm s}$\,$<$\,$-0.5$\,mag with TOPCAT. We also collected
  multi-wavelength photometry, constructed the spectral energy
  distributions and estimated effective temperatures from fits to
  atmospheric models with VOSA for the most interesting targets.}
{We assembled a sample of 32 bright blue high proper motion objects,
  including ten sdO/B subdwarfs, nine DA white dwarfs, five young
  early-type stars (two of which are runaway stars), two blue
  horizontal branch stars, one star with poor information, and five
  objects reported for the first time in this work. These last five
  objects have magnitudes $B_T$\,$\approx$\,11.0\,--\,11.6\,mag,
  effective temperatures $T_{\rm
    eff}$\,$\approx$\,24,000\,--\,30,000\,K, and are located in the
  region of known white dwarfs and hot subdwarfs in a reduced proper
  motion-colour diagram. We confirmed the hot subdwarf nature of one
  of the new objects, Albus\,5, with public far-ultraviolet
  spectroscopic data obtained with {\em FUSE}.}
{}
\keywords{astronomical data bases: miscellaneous -- virtual
  observatory tools -- stars: early-type -- stars: peculiar --
  subdwarfs -- white dwarfs}
\maketitle
%

\section{Introduction}
\label{introduction}

Bright objects with blue colours and high proper motions are rare in
the sky. By ``bright'' we mean sources with optical magnitudes in the
range covered by classical astronomical catalogues, such as the Bonner
Durchmusterung \citep{Schonfeld1886,Argelander03}, C\'ordoba
Durchmusterung \citep{Thome1894}, Henry Draper \citep[][and previous
  versions]{Cannon25}, or the most recent {\em Hipparcos}
\citep{Perryman97a} and Tycho-2 \citep{Hog00a} catalogues. These
catalogues typically have higher limit magnitudes between 9 and
11\,mag in the visible.  Blue colours are indicative of high effective
temperatures, while high proper motions of tens of mas\,yr$^{-1}$
imply short heliocentric distances or large tangential velocities at
moderate distances.

There are few possible kinds of bright blue high proper motion
objects. They can be nearby white dwarfs, hot subdwarfs, runaway OB
stars, or early-type stars in nearby young moving groups. Most of the
known blue objects (e.g., massive early-type stars in massive
clusters) in the Galaxy are located far away and, therefore, have low
proper motions and are apparently faint save for a few
exceptions. Likewise, a wealth of objects with high proper motions are
known, but the bulk of them are neighbouring late-type stars with red
colours. Blue extragalactic objects are also known, but they present a
negligible proper motion.

Bright blue high proper motion objects are important for many fields
in astrophysics. Bright white dwarfs and hot subdwarfs, for instance,
are extensively used as spectrophotometric standard stars
\citep{Stone77,Oke83,Oke90}, class prototypes, or as tracers of the
Population~II in the Galaxy. Their apparent high optical brightnesses
facilitate the determination of physical parameters at extreme
conditions (very high temperatures and gravities) in a relatively
undemanding way. Similarly, runaway OB stars and early-type stars in
young moving groups are fundamental for understanding the evolution
and evaporation of star-forming regions.

In this paper, we used Tycho-2 and the Two Micron All Sky Survey Point
Source (2MASS; \citealt{Skrutskie06}) catalogues and Virtual
Observatory tools to look for the bluest objects ($V_T-K_{\rm
  s}$\,$\leq$\,$-0.5$\,mag) with the highest proper motions
($\mu$\,$>$\,50\,mas\,yr$^{-1}$).

\section{Analysis and results}
\label{analysisandresults}

\subsection{Tycho-2/2MASS cross-match}

On the one hand, Tycho-2 gives position and proper motion information
for the 2.5 million brightest stars in the sky. Photometric data for
two pass-bands ({\it B$_T$} and {\it V$_T$}, close to Johnson B and V;
\citealt{Perryman97a}) are also provided. Typical uncertainties are
60\,mas in position, 2.5\,mas\,yr$^{-1}$ in proper motion, and
0.1\,mag in photometry. On the other hand, 2MASS contains photometric
data for 471 million sources in the $J$, $H$, and $K_{\rm s}$
near-infrared bands.  Typical photometric and astrometric
uncertainties of 2MASS are less than 0.03\,mag and 100\,mas,
respectively.

In this work we took advantage of Virtual Observatory\footnote{\tt
  http://www.ivoa.net} (VO) tools to cross-match the whole Tycho-2 and
2MASS catalogues. The VO is a project designed to provide the
astronomical community with the data access and the research tools
necessary to enable the exploration of the digital, multi-wavelength
universe resident in the astronomical data archives. In particular, we
made use of the scripting capability of Aladin\footnote{\tt
  http://aladin.u-strasbg.fr/} \cite[][]{Bonnarel00} to perform our
cross-match, a VO-compliant software that allows users to visualise
and analyse digitised astronomical images, and superimpose entries
from astronomical catalogues or databases available from the VO
services.  To avoid memory overflow during data processing, we divided
the full sky sphere into overlapping circular regions of 4.5\,deg$^2$.
The useful and powerful Aladin script mode allowed us to readily
perform the following workflow for each region:

\begin{figure}
\centering
\includegraphics[width=0.49\textwidth]{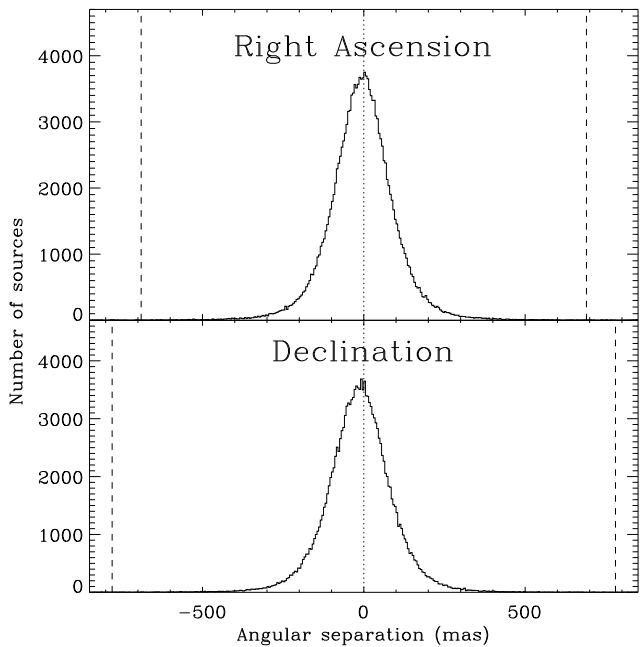}
\caption{Histograms of the angular separation in right ascension
  (upper panel) and declination (lower panel) between the Tycho-2 and
  the closest 2MASS counterpart, after correcting the 2MASS
  coordinates for Tycho-2 proper motion. The vertical dashed lines
  indicate the maximum separation allowed for a reliable cross-match,
  3\,$\sigma_{\rho_{\alpha}}$\,=\,0.69\,arcsec in right ascension, and
  3\,$\sigma_{\rho_{\delta}}$\,=\,0.78\,arcsec in declination.}
\label{DistH}
\end{figure}

\begin{itemize}

   \item Select all Tycho-2 sources with proper motion
     $\mu$\,$>$\,50\,mas\,yr$^{-1}$.

   \item Select all 2MASS sources within a radius of 40\,arcsec around
     each selected high proper motion Tycho-2 source.

   \item Convert 2MASS coordinates into J2000.0 epoch using the proper
     motion of the related Tycho-2 source.

   \item Select for each Tycho-2 source the nearest proper
     motion-corrected 2MASS source.
     
\end{itemize}

In the second step, we used such a large radius to account for the
expected high proper motion of the objects along the years. We took as
reference the maximum proper motion known so far (that of the
\object{Barnard star}, 10.3\,arcsec\,yr$^{-1}$) and the maximum
possible temporal separation between Tycho-2 and 2MASS observations
($\approx$\,4\,yr).

During the cross-match process, we imposed a good quality flag (Qflg =
A, B, C, or D) in $K_{\rm s}$ and an error lower than 0.3\,mag in
$V_T$. It returned more than 300,000 identifications. However, because
we worked with overlapping regions, an important fraction of the
identifications were duplicated. We used TOPCAT\footnote{\tt
  http://www.star.bris.ac.uk/$\sim$mbt/topcat/} to purge the
duplicated data and to carry out the subsequent analyses. TOPCAT is an
interactive graphical viewer and editor for tabular data that allows
the user to examine, analyse, combine, and edit astronomical
tables. After purging the data we ended with a sample of 157,184
sources fulfilling all the above criteria.

In Fig.~\ref{DistH} we show the histograms of angular separations
$\rho$ between the cross-matched Tycho-2 and 2MASS proper
motion-corrected sources, in both directions, right ascension and
declination. Most of the separations (98.3\,\%) are lower than
0.50\,arcsec. Quantitatively, the mean angular separation and standard
deviation are $\overline{\rho_{\alpha}}$\,=\,0.00\,arcsec and
$\sigma_{\rho_{\alpha}}$\,=\,0.23\,arcsec for the right ascension, and
$\overline{\rho_{\delta}}$\,=\,0.01\,arcsec and
$\sigma_{\rho_{\delta}}$\,=\,0.26\,arcsec for the declination. Thus,
we discarded 1,800 (spurious) matches with separations higher than
$3\sigma_\rho$ in any of the spatial directions, which reduced the
total number of sources to 155,384.

\subsection{Colour selection}

\begin{figure}
\centering
\includegraphics[width=0.49\textwidth]{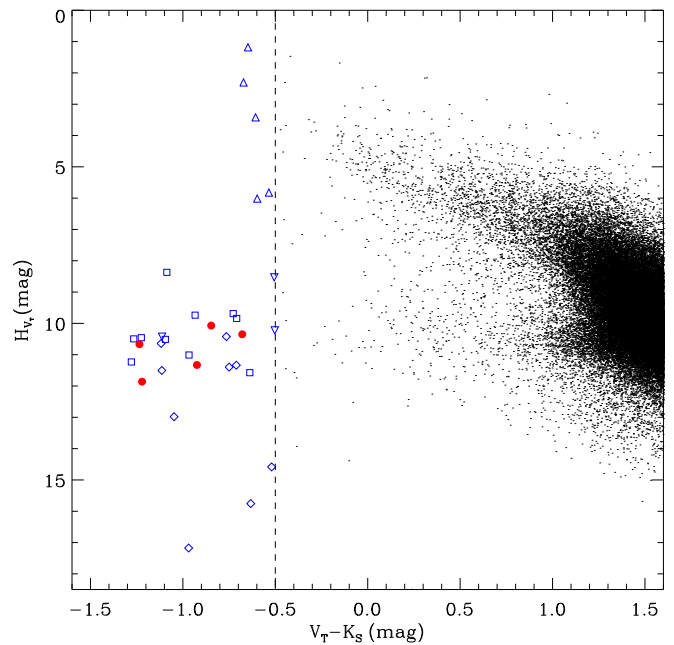}
\caption{Reduced proper motion diagram ($H_{V_T}$ versus
  $V_T-K_S$). Selected objects are bluewards of the dashed vertical
  line at $V_T-K_{\rm s}$\,=\,$-0.5$\,mag. Previously unreported blue
  high proper motion objects are depicted with (red) filled
  circles. Already known blue high proper motion objects are shown
  with (blue) open symbols: hot subdwarfs with squares, white dwarfs
  with diamonds, young main-sequence stars with up-triangles, and
  other stars with down-triangles. Objects redder than $V_T-K_{\rm
    s}$\,=\,$-0.5$\,mag are marked with (black) small dots.}
\label{rpm}
\end{figure}

We show the reduced proper motion-colour diagram of the cross-matched
sources in Fig.~\ref{rpm}. The reduced proper motion, $H_{V_T}$, is
defined as $H_{V_T}$\,=\,$V_T$\,+\,5\,$\log{\mu}$\,+\,5, where $\mu$
is the proper motion in arcseconds per year. It can also be expressed
as $H_{V_T}$\,=\,M$_{V_{T}}$\,+\,5\,$\log{\nu_t}$\,$-3.3$, where
$M_{V_{T}}$ is the $V_T$-band absolute magnitude and $\nu_t$ is the
transverse velocity in kilometers per second.  The $V_T-K_{\rm s}$
colour is a suitable effective-temperature indicator in field objects.
The reduced proper motion diagram therefore constitutes a powerful
tool for segregating members of kinematically-distinct stellar
populations \citep[e.g.][]{Lepine05}. In particular, it is possible to
distinguish three types of objects from Fig.~\ref{rpm}: Population~I
(upper branch), Population~II stars (middle branch) and white dwarfs
(very tiny lower branch).

For selecting blue high proper motion candidates, we applied a very
simple selection criterion: $V_T-K_{\rm s}$\,$<$\,$-0.5$\,mag, depicted
in Fig.~\ref{rpm} with a vertical dashed line. In total, 33 objects
showed bluer colour than this value. This simple selection criterion
produced an appropriate number of sources that could be studied in
detail in a reasonable time scale.

\subsection{Visual inspection, object rejection, and final sample}
\label{section.results}

\begin{figure}
\centering
\includegraphics[width=0.49\textwidth]{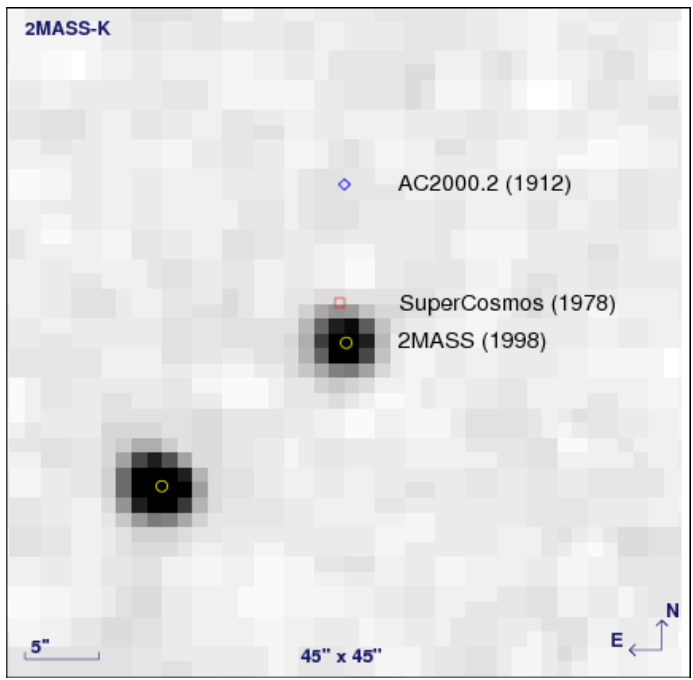}
\caption{Verification of the high proper motion of Albus~4
  (TYC\,6017--419--1), a representative blue high proper motion
  object, in a $K_{\rm s}$-band 2MASS image. The epochs of observation
  in the AC2000.2 (blue diamond), SuperCOSMOS (red square), and 2MASS
  (yellow circles) are shown in parenthesis. The DENIS epoch almost
  coincides with the 2MASS one.}
\label{albus4}
\end{figure}

We visually inspected the 33 selected blue high proper motion
candidates with Aladin. We compared and blinked two sets of images
separated several decades in time, optical DSS-1 and near-infrared
2MASS, to easily confirm the high proper motion of our candidates.
The Tycho-2 sources and 2MASS counterparts were superimposed on the
images to assess the reliability of the cross-match. Besides, we used
other available astro-photometric databases, such as the Astrographic
Catalogue AC2000.2 \citep{Urban98}, USNO-B1 \citep{Monet03},
SuperCOSMOS Sky Survey \citep{Hambly01a}, DENIS \citep{Epchtein97} and
the Positions and Proper Motions-Extended PPMX \citep{Roser08}, to
verify the high proper motion (Fig.~\ref{albus4}).

Of the 33 blue high proper motion candidates, we discarded
TYC\,8250--687--1, for which Tycho-2 provided an erroneous proper
motion after mixing the star 2MASS\,13052225--4740536
($V$\,$<$\,10\,mag) with the close ($\rho$\,$\approx$\,14\,arcmin)
bright extended {\em galaxy} \object{2MASX\,13052358--4740500}
($z$\,=\,0.01731\,$\pm$\,0.00015; \citealt{Crook07}).

\begin{table*}
  \caption[]{Already known bright blue high proper motion objects.}
  \label{known-BBHPM}
  \begin{tabular}{l cc c lll}
    \hline
    \hline
    \noalign{\smallskip}
TYC           	& $\alpha^{a}$ 	& $\delta^{a}$ 	& $\pi$$^{b}$ &  Name      		 & Object type 					      & Ref.$^{d}$ \\
               	& (J2000)     	& (J2000)      	&  (mas)                		 &             					      &           	\\
    \noalign{\smallskip}
    \hline
    \noalign{\smallskip}
6429--796--1	&  01 43 48.55  & --24 05 10.2  &   ~~3$\pm$4    & \object{CD--24 731}   & sdO/B hot subdwarf in a close binary system	      & 1,2,3 \\
8055--1270--1	&  02 16 30.58  & --51 30 43.8  & 21.22$\pm$0.12 & \object{$\phi$ Eri}	 & Young B8V star in Tucana-Horologium		      & 4 \\
9145--601--1	&  03 10 31.02  & --68 36 03.4  &  97.7$\pm$1.8  & \object{WD 0310--688} & DA3.1 white dwarf in a wide binary system	      & 5,6,7 \\
83--1181--1	&  04 44 42.13  &  +00 34 05.2  & ~~3.0$\pm$0.6  & \object{HD 30112}	 & Runaway B2.5V star				      & 8 \\
3734--2324--1  	&  05 05 30.61  &  +52 49 51.9  &    18$\pm$3    & \object{WD 0501+527}  & DA1 white dwarf  				      & 9,10,6 \\
9166--716--1   	&  05 31 40.35  & --69 53 02.2  &      ~~...     & \object{AA Dor}	 & sdO/B hot subdwarf in an eclipsing binary system   & 11,2,12 \\
7075--2143--1  	&  06 04 20.27  & --32 10 20.8  & ~~3.2$\pm$0.3  & \object{72 Col} AB	 & Runaway B2.5V star in a binary system	      & 13,14,8 \\
7613--1087--1  	&  06 23 12.20  & --37 41 29.0  &       ~~...    & \object{WD 0621--376} & DA1 white dwarf  				      & 15,10,6 \\
7664--1961--1  	&  08 10 31.60  & --40 33 10.0  &       ~~...    & \object{CD--40 3927}  & sdO hot subdwarf 				      & 2 \\
4547--1009--1  	&  09 21 19.18  &  +81 43 27.6  &  ~~4$\pm$2     & \object{AG+81 266}	 & sdO hot subdwarf 				      & 16,2 \\
3429--1180--1  	&  09 30 46.78  &  +48 16 23.8  &~~6.2$\pm$1.8   & \object{BD+48 1777}	 & sdOp hot subdwarf			       	      & 17,18,2 \\
3005--579--1   	&  10 35 16.71  &  +40 21 13.6  &       ~~...    & \object{PB 385} 	 & sdB hot subdwarf 				      & 19,2 \\
1984--97--1    	&  11 37 05.10  &  +29 47 58.3  &    63$\pm$4    & \object{WD 1134+300}  & DA2.5 white dwarf  				      & 20,6,7 \\
2541--4--1     	&  13 23 35.26  &  +36 07 59.5  &       ~~...    & \object{NSV 19768}	 & sdOp hot subdwarf 			       	      & 21,22,2 \\
5545--1390--1  	&  13 30 13.64  & --08 34 29.5  &    58$\pm$4    & \object{WD 1327--083} & DA3.7 white dwarf in a wide binary system  	      & 23,6,7 \\
2004--242--1   	&  13 38 24.76  &  +29 21 56.0  & ~~5.5$\pm$1.5  & \object{BD+30 2431}	 & B3Vp halo blue horizontal branch star	      & 24,25,26 \\ 
7832--2490--1  	&  14 58 31.93  & --43 08 02.3  &~~8.52$\pm$0.18 & \object{$\beta$ Lup}	 & Young B2III star in Sco OB2			      & 27,28 \\
7839--874--1   	&  15 26 07.13  & --39 29 19.2  &       ~~...    & \object{CD--39 9849}  & sdO/B hot subdwarf$^{c}$                           & 29,30 \\ 
7853--559--1   	&  16 23 33.84  & --39 13 46.2  &    76$\pm$3    & \object{WD 1620--391} & DA2 white dwarf in a wide binary system  	      & 31,6,7 \\
3885--860--1   	&  16 48 25.64  &  +59 03 22.7  &    94$\pm$3    & \object{DN Dra}	 & DAV4 white dwarf, pulsating 			      & 32,10,6 \\
8359--3650--1  	&  18 26 58.42  & --45 58 06.5  & 11.74$\pm$0.17 & \object{$\alpha$ Tel} & Young B3IV star in Sco OB2			      & 33,28 \\
3531--1327--1  	&  18 34 08.90  &  +48 27 17.0  &       ~~...    & \object{BD+48 2721}	 & sdB hot subdwarf 				      & 34,2,35 \\
2161--1038--1  	&  20 34 21.88  &  +25 03 49.7  &    64$\pm$3    & \object{WD 2032+248}  & DA2.5 white dwarf  				      & 23,6,7 \\
8424--1008--1  	&  21 39 10.61  & --46 05 51.5  & ~~3.8$\pm$1.7  & \object{HD 205805}	 & sdB hot subdwarf, helium-poor		      & 36,2 \\ 
2211--1613--1  	&  21 59 41.97  &  +26 25 57.4  & ~~9.1$\pm$1.2  & \object{IS Peg}	 & sdO hot subdwarf in a binary system		      & 37,2 \\
8441--1261--1  	&  22 14 11.91  & --49 19 27.3  &       ~~...    & \object{WD 2211--495} & DA1 white dwarf				      & 15,10,6 \\
6401--39--1    	&  23 12 30.63  & --21 06 43.4  & ~~0.0$\pm$1.4  & \object{HD 218970}	 & B2--3IV/Vp star found far from the galactic plane  & 18,38,39 \\
    \noalign{\smallskip}
           \hline
  \end{tabular}
\begin{list}{}{}
\item[$^{a}$] Equatorial coordinates from Tycho-2.
\item[$^{b}$] Trigonometric parallax from {\em Hipparcos}.
\item[$^{c}$] CD--39~9849 was classified as a blue horizontal branch
  star or sdO/B hot subdwarf by \cite{Drilling95}. We classify it as
  an sdO/B hot subdwarf based on its location in the reduced proper
  motion-colour diagram.
\item[$^{d}$] References.
  (1)\,\citealt{Haro62};
  (2)\,\citealt{Kilkenny88};
  (3)\,\citealt{O'Toole06};
  (4)\,\citealt{Zuckerman01};
  (5)\,\citealt{Thackeray61};
  (6)\,\citealt{Caballero07};
  (7)\,\citealt{Sion09};
  (8)\,\citealt{Hoogerwerf01};
  (9)\,\citealt{Greenstein69};
  (10)\,\citealt{McCook99};
  (11)\,\citealt{Luyten57};
  (12)\,\citealt{Rucinski09};
  (13)\,\citealt{Voute18};
  (14)\,\citealt{vanAlbada61};
  (15)\,\citealt{Holberg93};
  (16)\,\citealt{Berger78};
  (17)\,\citealt{Giclas67};
  (18)\,\citealt{Greenstein74};
  (19)\,\citealt{Berger77};
  (20)\,\citealt{Giclas65};
  (21)\,\citealt{Humason47};
  (22)\,\citealt{Greenstein66};
  (23)\,\citealt{Luyten49};
  (24)\,\citealt{Feige58};
  (25)\,\citealt{Baschek76};
  (26)\,\citealt{Bonifacio95};
  (27)\,\citealt{Pickering1897};
  (28)\,\citealt{Brown97};
  (29)\,\citealt{Drilling95};
  (30)\,{\em this work};
  (31)\,\citealt{Stephenson68};
  (32)\,\citealt{Schwartz72};
  (33)\,\citealt{Kapteyn14};
  (34)\,\citealt{Carnochan83};
  (35)\,\citealt{Edelmann01};
  (36)\,\citealt{Baschek70};
  (37)\,\citealt{Gould57}; 
  (38)\,\citealt{Conlon90};
  (39)\,\citealt{Martin06}.
\end{list}
\end{table*}

Of the 32 remaining objects, 27 were already known. We list their
basic properties in Table~\ref{known-BBHPM}. Among them are most of
the known brightest white dwarfs and hot subdwarfs. Besides, there are
runaway OB stars, early-type stars in young moving groups, and one
chemically-peculiar halo blue horizontal branch star. Several of them
have their trigonometric parallaxes measured by the {\em Hipparcos}
satellite \citep{vanLeeuwen07}. The trigonometric parallaxes of white
dwarfs are larger than those of hot subdwarfs, which indicates longer
heliocentric distances and higher luminosities for the latter,
consistent with their classification.

The remaining five objects, listed in Table~\ref{unknown-BBPM}, have never
before been reported in the literature. For naming them, we followed
the ``Albus'' nomenclature introduced by \citet{Caballero07} and
followed in \citet{Caballero09}. Our new Albus objects go from the
fourth to the eighth of this series.

\section{Discussion}
\label{discussion}

\begin{table*}
  \caption[]{New bright blue high proper motion objects.}
  \label{unknown-BBPM}
  \begin{tabular}{l ccccc}
    \hline
    \hline
    \noalign{\smallskip}
    Albus 					& 4 			& 5            		& 6    			& 7             	& 8            \\
    \noalign{\smallskip}
    \hline
    \noalign{\smallskip}
    TYC                                      	& 6017--419--1		& 4406--285--1        	& 9044--1653--1  	& 9327--1311--1    	& 4000--216--1 \\
    $\alpha^{a}$ (J2000)                     	& 08 51 58.89      	& 14 21 27.86         	& 16 00 11.81    	& 21 53 41.26      	& 23 34 52.08 \\
    $\delta^{a}$ (J2000)                     	& --17 12 38.8     	& +71 24 21.3         	& --64 33 30.3   	& --70 04 31.5     	& +53 47 02.5 \\
    $\mu_\alpha \cos{\delta}$ [mas\,a$^{-1}$] 	& +0$\pm$2         	& --62$\pm$2        	& +3$\pm$3~      	& +10$\pm$2~~      	& +68$\pm$2~~ \\
    $\mu_\delta$ [mas\,a$^{-1}$]              	& --124$\pm$2	   	& --15$\pm$2        	& --78$\pm$3   		& --50$\pm$2~~     	& +6$\pm$2~ \\
    $H_{V_{T}}$ [mag]                         	& 11.86$\pm$0.11   	& 10.35$\pm$0.12      	& 11.3$\pm$0.2   	& 10.07$\pm$0.16   	& 10.67$\pm$0.12 \\
    \noalign{\smallskip}
    Tycho $B_T$ [mag]                      	& 11.56$\pm$0.07   	& 10.97$\pm$0.04      	& 11.58$\pm$0.07 	& 11.28$\pm$0.05   	& 11.53$\pm$0.06 \\
    Tycho $V_T$ [mag]                      	& 11.40$\pm$0.09   	& 11.34$\pm$0.08      	& 11.86$\pm$0.15 	& 11.53$\pm$0.09   	& 11.49$\pm$0.09 \\
    DENIS $i$ [mag]                        	& 12.06$\pm$0.03   	& ...                   & 12.23$\pm$0.04 	& 11.83$\pm$0.06   	& ... \\
    2MASS $J$ [mag]                        	& 12.41$\pm$0.02   	& 11.85$\pm$0.02      	& 12.57$\pm$0.03 	& 12.15$\pm$0.02   	& 12.50$\pm$0.03 \\
    2MASS $H$ [mag]                        	& 12.55$\pm$0.02   	& 11.95$\pm$0.02$^{b}$ 	& 12.73$\pm$0.04 	& 12.26$\pm$0.03   	& 12.66$\pm$0.03  \\
    2MASS $K_s$ [mag]                      	& 12.62$\pm$0.03   	& 12.02$\pm$0.03       	& 12.77$\pm$0.04 	& 12.38$\pm$0.03   	& 12.73$\pm$0.02 \\
    \noalign{\smallskip}
    GRIZ $g$ [mag]                         	& 11.51            	& 11.17                	& ...               	& 11.48            	& 11.57 \\
    GRIZ $r$ [mag]                         	& 12.03            	& 11.64                	& ...                	& 11.95            	& 12.13 \\
    GRIZ $i$ [mag]                         	& 12.37            	& 11.94                	& ...                	& 12.26            	& 12.47 \\
    GRIZ $z$ [mag]                         	& 12.67            	& 12.21                	& ...                	& 12.52            	& 12.82 \\
    \noalign{\smallskip}
    $T_{\rm eff}^{c}$ [K]                  	& 27,000           	& 24,000        	& 29,000         	& 24,000           	& 30,000 \\ 
  \noalign{\smallskip}
  \hline
  \end{tabular}
  \begin{list}{}{}
     \item[$^{a}$] Equatorial coordinates from Tycho-2.
     \item[$^{b}$] Poor $H$-band 2MASS photometry quality (flag ``E'').
     \item[$^{c}$] Effective temperatures from the SED fits.
  \end{list}
\end{table*}

The five new bright blue high proper motion objects
(Table~\ref{unknown-BBPM}) are placed in the reduced proper
motion-colour diagram (Fig.~\ref{rpm}) in the region at which hot
subdwarfs and early-type white dwarfs gather. To shed some light into
the nature of the new Albus objects, we searched for additional
photometric data with the ``all-VO'' utility of Aladin. This utility
allows the user to query a large number of photometric catalogues in a
comfortable way. Besides Tycho-2 and 2MASS, two others catalogues had
additional photometric data: DENIS and GRIZ \citep{Ofek08}. DENIS
gives magnitudes in the $i$ band for the three Albus sources in the
Southern hemisphere, whereas the GRIZ catalogue provides synthetic
SDSS $griz$ magnitudes for all new Albus sources but Albus\,6 (see
Table~\ref{unknown-BBPM}). The typical uncertainty for the GRIZ
photometry is about 0.12, 0.12, 0.10, and 0.08 mag (1\,$\sigma$), for
the $g$, $r$, $i$, and $z$-bands, respectively.

\begin{figure}
\centering
\includegraphics[width=0.49\textwidth]{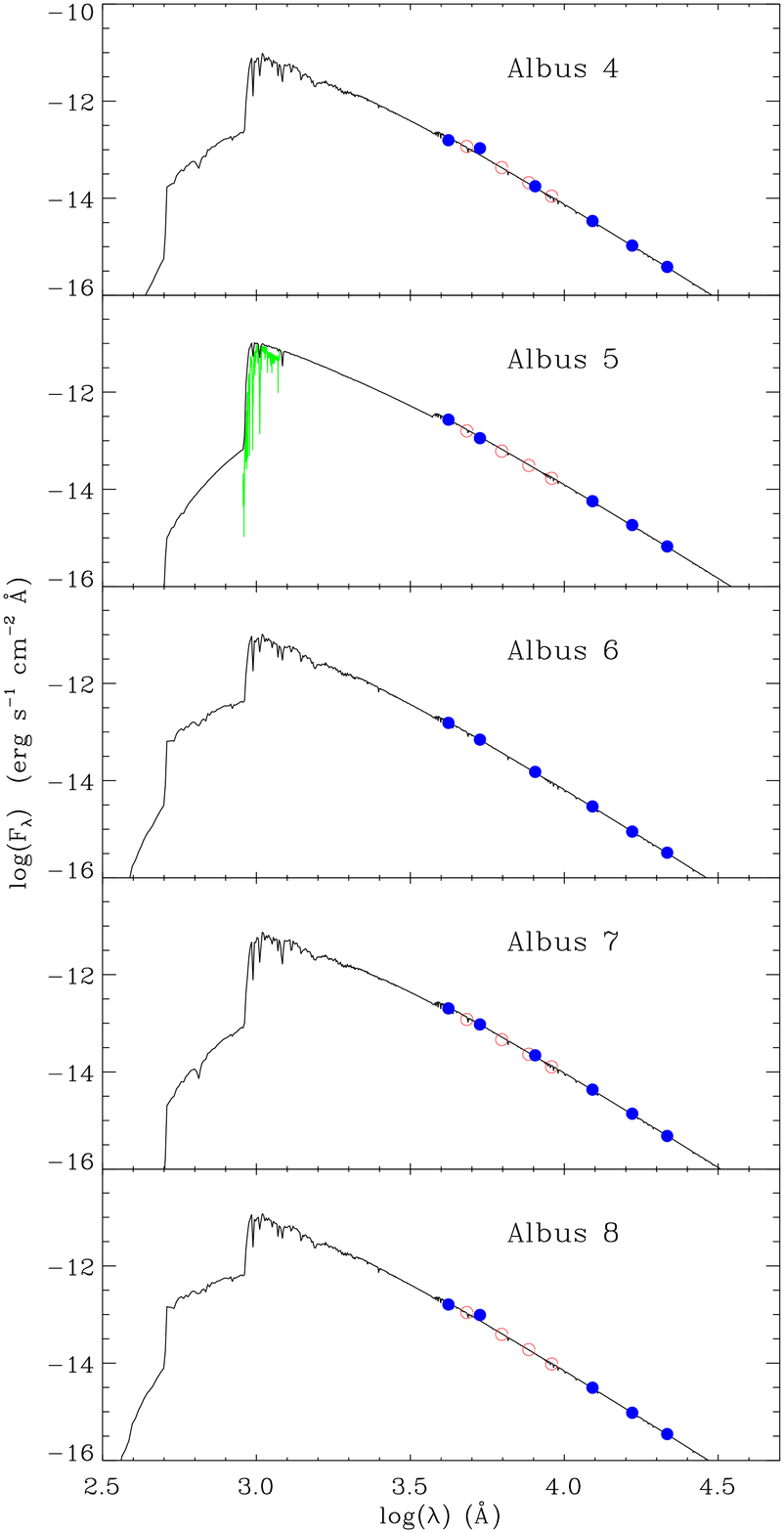}
\caption{Theoretical fitting of the Albus SEDs. Solid lines represent
  the Kurucz models that best fits the observational data. Filled
  (blue) and open (red) circles indicate the photometric data used for
  the fit and the synthetic GRIZ photometry data used for control,
  respectively. The thin (green) line in the second panel depicts the
  ultraviolet {\em FUSE} spectra of Albus\,5.}
\label{SEDs}
\end{figure}

We took advantage of another VO-tool, VOSA\footnote{\tt
  http://svo.cab.inta-csic.es/theory/vosa} (VO Sed Analyzer;
\citealt{Bayo08}) to fit the observed spectral energy distributions
(SEDs) to theoretical models. The VOSA allows the user to query in an
automatic and transparent way different collections of theoretical
models, to calculate their synthetic photometry, and to perform a
statistical test to determine which model best reproduces the observed
data. To fit the SEDs of our Albus objects, we used only the
observational data (all but the GRIZ ones) and fitted them to Kurucz
stellar atmosphere models \citep{Castelli97}. The effective
temperatures $T_{\rm eff}$ obtained with VOSA ranged between 24,000
and 30,000\,K (see Table~\ref{unknown-BBPM}). The accuracy of the
effective temperature estimated with this method is defined by the
step of the grid of the model, 1,000\,K in this case. In
Fig.\,\ref{SEDs} we have plotted the theoretical fitting of the
observational SEDs.

\begin{figure}
\centering
\includegraphics[width=0.49\textwidth]{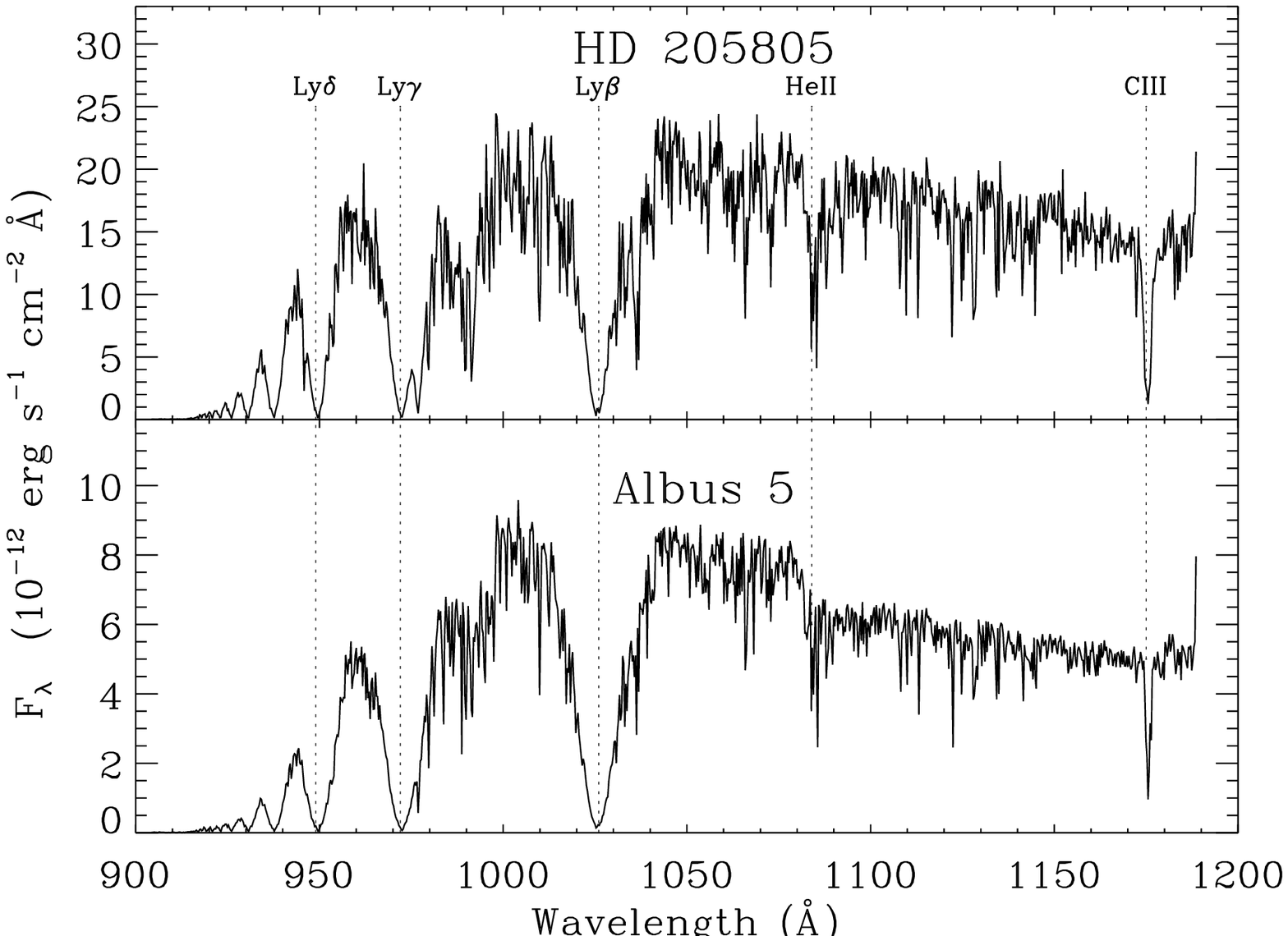}
\caption{{\em FUSE} spectra of the sdB star HD\,205805 (upper panel)
  and Albus\,5 (lower panel). The strongest observed lines (e.g. of
  the Lyman series) and multiplets (e.g. C\,{\sc iii}) have been
  labelled.}
\label{fuse}
\end{figure}

The confirmation of the real nature of these objects requires
spectroscopic data. To look for spectra in the VO archives, we used
another VO-tool, VOSED\footnote{\tt
  http://sdc.cab.inta-csic.es/vosed}. Developed by the Spanish Virtual
Observatory, VOSED allows the user to gather spectroscopic information
available throughout the VO. Only Albus\,5 was found to have
spectroscopic data. In particular, it was observed by the {\it Far
  Ultraviolet Spectroscopic Explorer} ({\em FUSE}) as part of the
programme G061. We used this spectrum to confirm its hot subdwarf
nature. In Fig.~\ref{fuse} we compare the {\em FUSE} spectra of
Albus\,5 to the well known sdB star \object{HD~205805} ($T_{\rm
  eff}$\,$\gtrsim$\,26\,000\,K; \citealt{Baschek70}). The two spectra
are almost identical, showing a huge number of photospheric and
interstellar absorption lines that are observed in the {\em FUSE}
spectra of almost all sdB stars \citep{Chayer06}. Because of the
strong similitude of these two ultraviolet spectra and the $T_{\rm
  eff}$ of 24,000\,K obtained using VOSA, we classify Albus\,5 as a
bona-fide sdB star. The other four objects have equal or higher $T_{\rm
  eff}$ (and bluer $V_T-K_{\rm s}$ colours) and could be either sdO/B
subdwarfs or early-type white dwarfs. Knowing their trigonometric
parallaxes would help to classify them. Unfortunately, neither of
the five new Albus are included in the {\em Hipparcos} catalogue.

\section{Summary}

To find out blue high proper motion objects, we cross-matched the 2.5
million sources in the Tycho-2 catalogue with the 471 million sources
in the 2MASS catalogue. From this correlation, and after discarding
one erroneous identification, we obtained a sample of 32 objects with
proper motions larger than 50\,mas\,yr$^{-1}$ and $V_T-K_{\rm s}$
colours bluer than $-0.5$\,mag. Of these, 27 were known and five
(Albus~4 to~8) were studied here for the first time.

From the position of the new discovered sources in a reduced proper
motion-colour diagram and theoretical fitting of their SEDs, we
concluded that these objects are probably hot subdwarfs or early-type
white dwarfs. For Albus\,5, we confirmed its nature as an sdB star
with a {\em FUSE} far-ultraviolet spectrum. For the remaining objects,
spectroscopic follow-up observations are required to ascertain their
true nature.  In any case, because of their relative brightness and
SED profiles, the five objects can serve as useful spectro-photometric
standards or be the subject of dedicated studies (e.g., pulsations and
origin of hot subdwarfs, metallic abundances and atmospheric
parameters in white dwarfs).

\begin{acknowledgements}
This work was partially funded by the Spanish Ministerio de Ciencia e
Innovaci\'on under the Consolider-Ingenio 2010 Program grant
CSD2006-00070 {\em First Science with the Gran Telescopio Canarias}.
This research has made use of VOSA and VOSED, developed by the Spanish
Virtual Observatory through grants AyA2008-02156 and RI031675, and of
Aladin and SIMBAD developed at the Centre de Donn\'ees astronomiques
de Strasbourg, France. Financial support was provided by the
Universidad Complutense de Madrid, the Comunidad Aut\'onoma de Madrid,
and the Spanish Ministerio de Ciencia e Innovaci\'on under grants
AyA2008-00695, 
AyA2008-06423-C03-03, 
SP2009/ESP-1496. 
\end{acknowledgements}

\bibliographystyle{aa} 
\bibliography{jimenez2010A+A525_A29} 

\begin{thebibliography}{62}
\expandafter\ifx\csname natexlab\endcsname\relax\def\natexlab#1{#1}\fi

\bibitem[{{Argelander}(1903)}]{Argelander03}
{Argelander}, F. 1903, {Bonner Durchmusterung des nordlichen Himmels. Eds
  Marcus and Weber's Verlag, Bonn}

\bibitem[{{Baschek} \& {Norris}(1970)}]{Baschek70}
{Baschek}, B. \& {Norris}, J. 1970, \apjs, 19, 327

\bibitem[{{Baschek} \& {Sargent}(1976)}]{Baschek76}
{Baschek}, B. \& {Sargent}, A.~I. 1976, \aap, 53, 47

\bibitem[{{Bayo} {et~al.}(2008){Bayo}, {Rodrigo}, {Barrado Y Navascu{\'e}s},
  {Solano}, {Guti{\'e}rrez}, {Morales-Calder{\'o}n}, \& {Allard}}]{Bayo08}
{Bayo}, A., {Rodrigo}, C., {Barrado Y Navascu{\'e}s}, D., {et~al.} 2008, \aap,
  492, 277

\bibitem[{{Berger} \& {Fringant}(1977)}]{Berger77}
{Berger}, J. \& {Fringant}, A. 1977, \aaps, 28, 123

\bibitem[{{Berger} \& {Fringant}(1978)}]{Berger78}
{Berger}, J. \& {Fringant}, A. 1978, \aap, 64, L9

\bibitem[{{Bonifacio} {et~al.}(1995){Bonifacio}, {Castelli}, \&
  {Hack}}]{Bonifacio95}
{Bonifacio}, P., {Castelli}, F., \& {Hack}, M. 1995, \aaps, 110, 441

\bibitem[{{Bonnarel} {et~al.}(2000){Bonnarel}, {Fernique}, {Bienaym{\'e}},
  {Egret}, {Genova}, {Louys}, {Ochsenbein}, {Wenger}, \&
  {Bartlett}}]{Bonnarel00}
{Bonnarel}, F., {Fernique}, P., {Bienaym{\'e}}, O., {et~al.} 2000, \aaps, 143,
  33

\bibitem[{{Brown} \& {Verschueren}(1997)}]{Brown97}
{Brown}, A.~G.~A. \& {Verschueren}, W. 1997, \aap, 319, 811

\bibitem[{{Caballero}(2009)}]{Caballero09}
{Caballero}, J.~A. 2009, {Multi-wavelength Astronomy and Virtual Observatory,
  Proceedings of the EURO-VO Workshop, held at the European Space Astronomy
  Centre of ESA, Villafranca del Castillo, Spain, 1-3 December, 2008, eds. D.
  Baines and P. Osuna, published by the European Space Agency, p.3}

\bibitem[{{Caballero} \& {Solano}(2007)}]{Caballero07}
{Caballero}, J.~A. \& {Solano}, E. 2007, \apjl, 665, L151

\bibitem[{{Cannon} \& {Pickering}(1925)}]{Cannon25}
{Cannon}, A.~J. \& {Pickering}, E. 1925, {The Henry Draper (HD) Catalogue.
  Vol.100: HD extension}, ed. {Cannon, A.~J.~\& Pickering, E.}

\bibitem[{{Carnochan} \& {Wilson}(1983)}]{Carnochan83}
{Carnochan}, D.~J. \& {Wilson}, R. 1983, \mnras, 202, 317

\bibitem[{{Castelli} {et~al.}(1997){Castelli}, {Gratton}, \&
  {Kurucz}}]{Castelli97}
{Castelli}, F., {Gratton}, R.~G., \& {Kurucz}, R.~L. 1997, \aap, 318, 841

\bibitem[{{Chayer} {et~al.}(2006){Chayer}, {Oliveira}, {Dupuis}, {Moos}, \&
  {Welsh}}]{Chayer06}
{Chayer}, P., {Oliveira}, C., {Dupuis}, J., {Moos}, H.~W., \& {Welsh}, B.~Y.
  2006, in Astronomical Society of the Pacific Conference Series, Vol. 348,
  Astrophysics in the Far Ultraviolet: Five Years of Discovery with FUSE, ed.
  {G.~Sonneborn, H.~W.~Moos, \& B.-G.~Andersson}, 209

\bibitem[{{Conlon} {et~al.}(1990){Conlon}, {Dufton}, {Keenan}, \&
  {Leonard}}]{Conlon90}
{Conlon}, E.~S., {Dufton}, P.~L., {Keenan}, F.~P., \& {Leonard}, P.~J.~T. 1990,
  \aap, 236, 357

\bibitem[{{Crook} {et~al.}(2007){Crook}, {Huchra}, {Martimbeau}, {Masters},
  {Jarrett}, \& {Macri}}]{Crook07}
{Crook}, A.~C., {Huchra}, J.~P., {Martimbeau}, N., {et~al.} 2007, \apj, 655,
  790

\bibitem[{{Drilling} \& {Bergeron}(1995)}]{Drilling95}
{Drilling}, J.~S. \& {Bergeron}, L.~E. 1995, \pasp, 107, 846

\bibitem[{{Edelmann} {et~al.}(2001){Edelmann}, {Heber}, \&
  {Napiwotzki}}]{Edelmann01}
{Edelmann}, H., {Heber}, U., \& {Napiwotzki}, R. 2001, AN, 322, 401

\bibitem[{{Epchtein} {et~al.}(1997){Epchtein}, {de Batz}, {Capoani},
  {Chevallier}, {Copet}, {Fouque}, {Lacombe}, {Le Bertre}, {Pau}, {Rouan},
  {Ruphy}, {Simon}, {Tiphene}, {Burton}, {Bertin}, {Deul}, {Habing},
  {Borsenberger}, {Dennefeld}, {Guglielmo}, {Loup}, {Mamon}, {Ng}, {Omont},
  {Provost}, {Renault}, {Tanguy}, {Kimeswenger}, {Kienel}, {Garzon}, {Persi},
  {Ferrari-Toniolo}, {Robin}, {Paturel}, {Vauglin}, {Forveille}, {Delfosse},
  {Hron}, {Schultheis}, {Appenzeller}, {Wagner}, {Balazs}, {Holl}, {Lepine},
  {Boscolo}, {Picazzio}, {Duc}, \& {Mennessier}}]{Epchtein97}
{Epchtein}, N., {de Batz}, B., {Capoani}, L., {et~al.} 1997, The Messenger, 87,
  27

\bibitem[{{Feige}(1958)}]{Feige58}
{Feige}, J. 1958, \apj, 128, 267

\bibitem[{{Giclas} {et~al.}(1965){Giclas}, {Burnham}, \& {Thomas}}]{Giclas65}
{Giclas}, H.~L., {Burnham}, R., \& {Thomas}, N.~G. 1965, Lowell Observatory
  Bulletin, 6, 155

\bibitem[{{Giclas} {et~al.}(1967){Giclas}, {Burnham}, \& {Thomas}}]{Giclas67}
{Giclas}, H.~L., {Burnham}, R., \& {Thomas}, N.~G. 1967, Lowell Observatory
  Bulletin, 7, 49

\bibitem[{{Gould} {et~al.}(1957){Gould}, {Herbig}, \& {Morgan}}]{Gould57}
{Gould}, N.~L., {Herbig}, G.~H., \& {Morgan}, W.~W. 1957, \pasp, 69, 242

\bibitem[{{Greenstein}(1966)}]{Greenstein66}
{Greenstein}, J.~L. 1966, \apj, 144, 496

\bibitem[{{Greenstein}(1969)}]{Greenstein69}
{Greenstein}, J.~L. 1969, \apj, 158, 281

\bibitem[{{Greenstein} \& {Sargent}(1974)}]{Greenstein74}
{Greenstein}, J.~L. \& {Sargent}, A.~I. 1974, \apjs, 28, 157

\bibitem[{{Hambly} {et~al.}(2001){Hambly}, {MacGillivray}, {Read}, {Tritton},
  {Thomson}, {Kelly}, {Morgan}, {Smith}, {Driver}, {Williamson}, {Parker},
  {Hawkins}, {Williams}, \& {Lawrence}}]{Hambly01a}
{Hambly}, N.~C., {MacGillivray}, H.~T., {Read}, M.~A., {et~al.} 2001, \mnras,
  326, 1279

\bibitem[{{Haro} \& {Luyten}(1962)}]{Haro62}
{Haro}, G. \& {Luyten}, W.~J. 1962, Boletin de los Observatorios Tonantzintla y
  Tacubaya, 3, 37

\bibitem[{{H{\o}g} {et~al.}(2000){H{\o}g}, {Fabricius}, {Makarov}, {Urban},
  {Corbin}, {Wycoff}, {Bastian}, {Schwekendiek}, \& {Wicenec}}]{Hog00a}
{H{\o}g}, E., {Fabricius}, C., {Makarov}, V.~V., {et~al.} 2000, \aap, 355, L27

\bibitem[{{Holberg} {et~al.}(1993){Holberg}, {Barstow}, {Buckley}, {Chen},
  {Dreizler}, {Marsh}, {O'Donoghue}, {Sion}, {Tweedy}, {Vauclair}, \&
  {Werner}}]{Holberg93}
{Holberg}, J.~B., {Barstow}, M.~A., {Buckley}, D.~A.~H., {et~al.} 1993, \apj,
  416, 806

\bibitem[{{Hoogerwerf} {et~al.}(2001){Hoogerwerf}, {de Bruijne}, \& {de
  Zeeuw}}]{Hoogerwerf01}
{Hoogerwerf}, R., {de Bruijne}, J.~H.~J., \& {de Zeeuw}, P.~T. 2001, \aap, 365,
  49

\bibitem[{{Humason} \& {Zwicky}(1947)}]{Humason47}
{Humason}, M.~L. \& {Zwicky}, F. 1947, \apj, 105, 85

\bibitem[{{Kapteyn}(1914)}]{Kapteyn14}
{Kapteyn}, J.~C. 1914, \apj, 40, 43

\bibitem[{{Kilkenny} {et~al.}(1988){Kilkenny}, {Heber}, \&
  {Drilling}}]{Kilkenny88}
{Kilkenny}, D., {Heber}, U., \& {Drilling}, J.~S. 1988, South African
  Astronomical Observatory Circular, 12, 1

\bibitem[{{L{\'e}pine} \& {Shara}(2005)}]{Lepine05}
{L{\'e}pine}, S. \& {Shara}, M.~M. 2005, \aj, 129, 1483

\bibitem[{{Luyten}(1949)}]{Luyten49}
{Luyten}, W.~J. 1949, \apj, 109, 528

\bibitem[{{Luyten}(1957)}]{Luyten57}
{Luyten}, W.~J. 1957, in The Observatory, Univ. Minnesota, Minneapolis, 1953,
  9, 1

\bibitem[{{Martin}(2006)}]{Martin06}
{Martin}, J.~C. 2006, \aj, 131, 3047

\bibitem[{{McCook} \& {Sion}(1999)}]{McCook99}
{McCook}, G.~P. \& {Sion}, E.~M. 1999, \apjs, 121, 1

\bibitem[{{Monet} {et~al.}(2003){Monet}, {Levine}, {Canzian}, {Ables}, {Bird},
  {Dahn}, {Guetter}, {Harris}, {Henden}, {Leggett}, {Levison}, {Luginbuhl},
  {Martini}, {Monet}, {Munn}, {Pier}, {Rhodes}, {Riepe}, {Sell}, {Stone},
  {Vrba}, {Walker}, {Westerhout}, {Brucato}, {Reid}, {Schoening}, {Hartley},
  {Read}, \& {Tritton}}]{Monet03}
{Monet}, D.~G., {Levine}, S.~E., {Canzian}, B., {et~al.} 2003, \aj, 125, 984

\bibitem[{{Ofek}(2008)}]{Ofek08}
{Ofek}, E.~O. 2008, \pasp, 120, 1128

\bibitem[{{Oke}(1990)}]{Oke90}
{Oke}, J.~B. 1990, \aj, 99, 1621

\bibitem[{{Oke} \& {Gunn}(1983)}]{Oke83}
{Oke}, J.~B. \& {Gunn}, J.~E. 1983, \apj, 266, 713

\bibitem[{{O'Toole} \& {Heber}(2006)}]{O'Toole06}
{O'Toole}, S.~J. \& {Heber}, U. 2006, \aap, 452, 579

\bibitem[{{Perryman} {et~al.}(1997){Perryman}, {Lindegren}, {Kovalevsky},
  {Hoeg}, {Bastian}, {Bernacca}, {Cr{\'e}z{\'e}}, {Donati}, {Grenon}, {van
  Leeuwen}, {van der Marel}, {Mignard}, {Murray}, {Le Poole}, {Schrijver},
  {Turon}, {Arenou}, {Froeschl{\'e}}, \& {Petersen}}]{Perryman97a}
{Perryman}, M.~A.~C., {Lindegren}, L., {Kovalevsky}, J., {et~al.} 1997, \aap,
  323, L49

\bibitem[{{Pickering} \& {Cannon}(1897)}]{Pickering1897}
{Pickering}, E.~C. \& {Cannon}, A.~J. 1897, \apj, 6, 349

\bibitem[{{R{\"o}ser} {et~al.}(2008){R{\"o}ser}, {Schilbach}, {Schwan},
  {Kharchenko}, {Piskunov}, \& {Scholz}}]{Roser08}
{R{\"o}ser}, S., {Schilbach}, E., {Schwan}, H., {et~al.} 2008, \aap, 488, 401

\bibitem[{{Rucinski}(2009)}]{Rucinski09}
{Rucinski}, S.~M. 2009, \mnras, 395, 2299

\bibitem[{{Schonfeld}(1886)}]{Schonfeld1886}
{Schonfeld}, E. 1886, in {\em Bonner Durchmusterung des nordlichen Himmels},
  eds. Marcus and Weber's Verlag, Bonn

\bibitem[{{Schwartz}(1972)}]{Schwartz72}
{Schwartz}, R.~D. 1972, \pasp, 84, 28

\bibitem[{{Sion} {et~al.}(2009){Sion}, {Holberg}, {Oswalt}, {McCook}, \&
  {Wasatonic}}]{Sion09}
{Sion}, E.~M., {Holberg}, J.~B., {Oswalt}, T.~D., {McCook}, G.~P., \&
  {Wasatonic}, R. 2009, \aj, 138, 1681

\bibitem[{{Skrutskie} {et~al.}(2006){Skrutskie}, {Cutri}, {Stiening},
  {Weinberg}, {Schneider}, {Carpenter}, {Beichman}, {Capps}, {Chester},
  {Elias}, {Huchra}, {Liebert}, {Lonsdale}, {Monet}, {Price}, {Seitzer},
  {Jarrett}, {Kirkpatrick}, {Gizis}, {Howard}, {Evans}, {Fowler}, {Fullmer},
  {Hurt}, {Light}, {Kopan}, {Marsh}, {McCallon}, {Tam}, {Van Dyk}, \&
  {Wheelock}}]{Skrutskie06}
{Skrutskie}, M.~F., {Cutri}, R.~M., {Stiening}, R., {et~al.} 2006, \aj, 131,
  1163

\bibitem[{{Stephenson} {et~al.}(1968){Stephenson}, {Sanduleak}, \&
  {Hoffleit}}]{Stephenson68}
{Stephenson}, C.~B., {Sanduleak}, N., \& {Hoffleit}, D. 1968, \pasp, 80, 92

\bibitem[{{Stone}(1977)}]{Stone77}
{Stone}, R.~P.~S. 1977, \apj, 218, 767

\bibitem[{{Thackeray}(1961)}]{Thackeray61}
{Thackeray}, A.~D. 1961, Monthly Notes of the Astronomical Society of South
  Africa, 20, 40

\bibitem[{{Thome}(1894)}]{Thome1894}
{Thome}, J.~M. 1894, {Cordoba Durchmusterung. Brightness and position of every
  fixed star down to the 10. magnitude comprised in the belt of the heavens
  between 32 and 90 degrees of southern declination - Vol.17: -32 deg. to -42
  deg.}, ed. J.~M. Thome

\bibitem[{{Urban} {et~al.}(1998){Urban}, {Corbin}, {Wycoff}, {Martin},
  {Jackson}, {Zacharias}, \& {Hall}}]{Urban98}
{Urban}, S.~E., {Corbin}, T.~E., {Wycoff}, G.~L., {et~al.} 1998, \aj, 115, 1212

\bibitem[{{van Albada}(1961)}]{vanAlbada61}
{van Albada}, T.~S. 1961, \bain, 15, 301

\bibitem[{{van Leeuwen}(2007)}]{vanLeeuwen07}
{van Leeuwen}, F. 2007, \aap, 474, 653

\bibitem[{{Vo{\^u}te}(1918)}]{Voute18}
{Vo{\^u}te}, J. 1918, \apj, 48, 144

\bibitem[{{Zuckerman} {et~al.}(2001){Zuckerman}, {Song}, \&
  {Webb}}]{Zuckerman01}
{Zuckerman}, B., {Song}, I., \& {Webb}, R.~A. 2001, \apj, 559, 388

\end{thebibliography}


\end{document}